# Property Graph Techniques in Relational Databases

Malcolm Crowe
University of the West of Scotland (retired)
Paisley, Scotland
e-mail: Malcolm.Crowe@uws.ac.uk

Fritz Laux
Reutlingen University (retired)
Reutlingen, Germany
e-mail: Friedrich.Laux@Reutlingen-University.de

*Abstract*— **Current developments in standardization activities for database management systems include the development of the next versions of the Structured Query Language (SQL), and related standards for labelled property graphs. This short paper explores mechanisms for unified implementation of record-based relational and property graph databases: its contribution is that modeling SQL foreign keys as reference keys facilitates the graph insert and advanced pattern matching mechanisms of Property Graph Database Management Systems (PGDMS) and allows the construction of multiple type graph models on a single relational database.**

## I. INTRODUCTION

Use cases where queries traverse many links have been shown to be inefficient in relational Database Management Systems (DBMS), because every link requires construction of a join [1]. Examples where queries must follow many links between data items include the investigation of financial fraud [2], supply chain and logistics analysis [3], protein-protein interactions in the cell [4], and social networks [5]. Efficient computation in such cases requires the construction of a suitable graph model, selecting the record types to form the nodes of a property graph, and then defining edges between these nodes [6]. Then queries can collect and analyze information resulting from traversals or pattern matching in the graph model [7].

Many products are already available that implement property graph database technology: The Graph Query Language (GQL) standard [8] v.1 emerged in 2024 [9], and the next version is expected in 2027. Most of these products stand alone, but Oracle's Structured Query Language/Property Graph Query (SQL/PGQ) offers a graph view over a relational database.

In this paper, we consider only record-based DBMS, where all durable data begins as a record laid down at a certain time (a persisted entity or object instance): it has an identity in addition to its values/attributes, may be referred to and possibly updated, and it becomes inaccessible if deleted. Many commercial DBMS including Oracle and DB2 have this model in the physical database, but it is not usually exposed to application programs. This paper presents a mechanism for high-performance graph management and pattern matching algorithms to be provided as an extension of any such relational DBMS, using the standard SQL reference type as a starting point.

The key contribution of this paper is that in moving from one node in a graph pattern to the next during a match statement we follow a reference or its reverse; and during a graph insert process, we build node references. We will show that reference values can replace foreign key indexes in implementations of both relational and graph databases.

Section II briefly reviews the reference type concept from SQL as a way of implementing foreign keys. It is shown that for a record-based DBMS, all foreign keys can be implemented using reference keys.

In Section III, we show that it is very straightforward to extend an ordinary relational DBMS to perform the "insert graph" and "match" operations that are key to property graph technology.

Section IV considers how alterative graph models can be supported by a relational database.

Section V presents the conclusions of this research, and a reference to an open-source implementation of these ideas.

## II. SQL AND REFERENCE VALUES

The SQL standard [9] offers few clues as to how the REF keyword should be implemented, and it is striking that the implementations in ADO.NET and Oracle are quite different. In ADO.NET we see that an entity reference has the form REF(expression) "where expression is any valid expression that yields an instance of an entity type". The entity reference consists of the entity key and an entity set name. In Oracle REF(correlation variable) "takes as its argument a correlation variable (table alias) associated with a row of an object table or view" and returns a REF value "for the object instance that is bound to the variable or row".

In both cases, it is easiest to think of reference values as belonging to a domain specific to an entity set or object table/view T. For convenience, let us call this domain REF T. It does not matter how the DBMS implements this, but it may be useful to think of it as a position in a real or conceptual log file. If the reference is a physical address a foreign key is replaced by an address (pointer) to the referenced record. This eliminates the need for a search (lookup) and leads directly to the referenced data record. It is easy to see that joins can benefit from such a substantial speed-up for large tables or multiple joins.

In the relational model, a foreign key in a relation R is a mapping from a set of one or more columns of R to the primary key of another table T. The primary key uniquely determines a record in T, and conversely if we know the reference to this record, we can read off the value of any of its columns, so that the foreign key can be equivalently represented as a mapping to REF T values. We can store this reference in the referencing table instead of storing the foreign key values themselves.

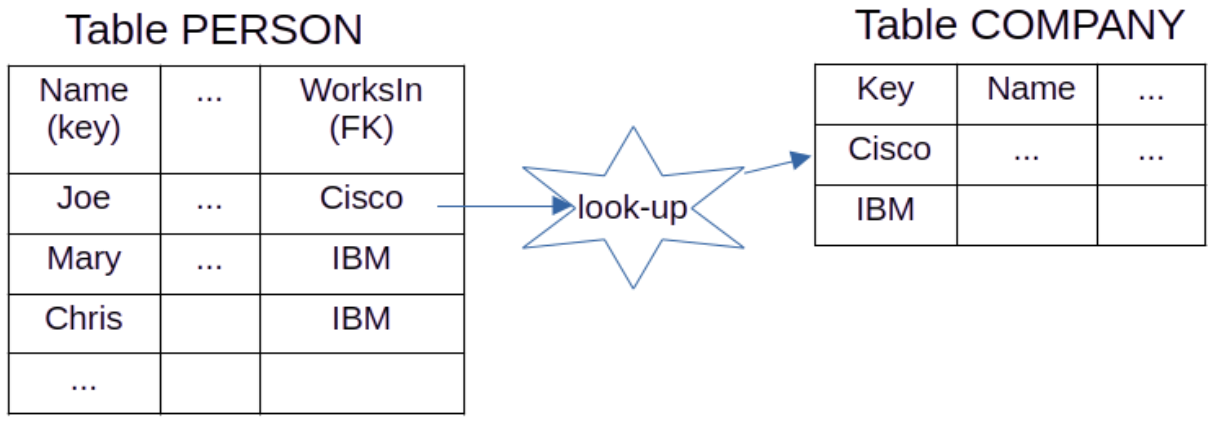

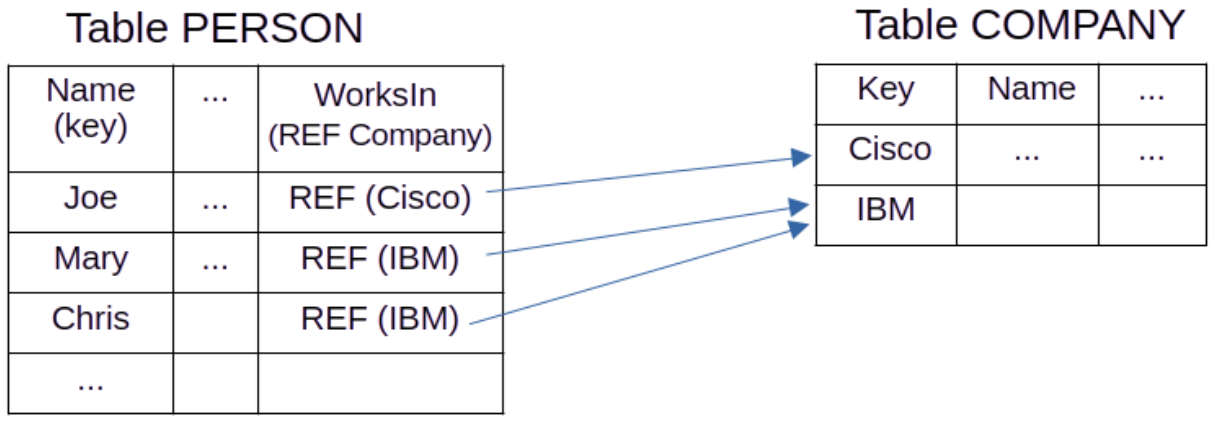

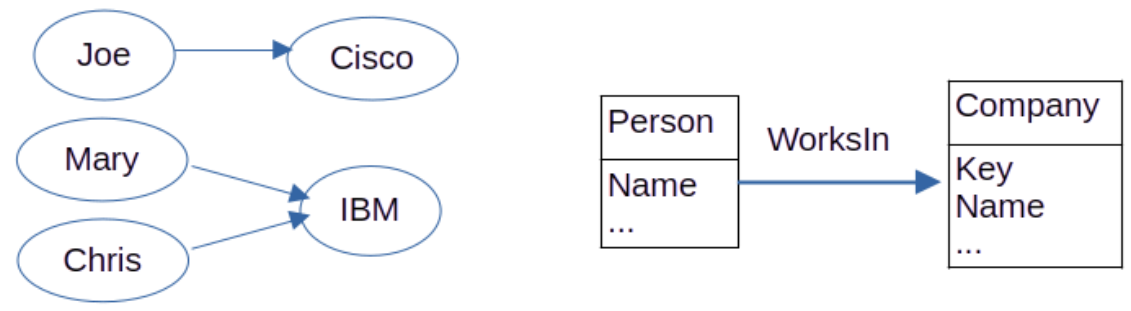

Figure 1. REF and foreign key values

For example, if we have a table called `Person` whose primary key column is name, a foreign key column `Worksin` might identify a record in a `Company` table.

```
Person (Name primary key,  …, WorksIn fkey(cid))
Company (cid primary key, Business, Town, … )
```

The Person table is a relation with one foreign key. This is a many-to-one relationship. Many-to-many relationships such as `Likes` need an extra table to represent them in SQL.

Such foreign keys correspond in graph databases to directional edges linking nodes representing rows in the `Person` table to rows in the `Company` table. In this example nodes would represent persons and companies, and directed edges would implement the `WorksIn` relationship.

But the SQL version of this graph model would have a separate table for `WorksIn` (the lookup table), so that each edge identifies a person and a company:

```
WorksIn (sid fkey (Name), .., cid fkey (Company))
```

Generally, edge relations have two references as here. In the ANSI/SPARC design method for databases, it was always the case that alternative data models could be used with a single database, and we return to this point later.

In this paper, let us notate by T(n) a reference value to the record in table T identified by the expression n, or REF(n) if the relevant table is clear from the context: where n might be a primary key. Figure 1 illustrates the above example where part a) shows the known relational case, where the foreign key value `'Cisco'` is mapped to a row in `Company`, part b) uses REF values, so that the `WorksIn` column contains a physical pointer to (or row uid of) a row in `Company`, and part c) shows the corresponding property graph model, where each instance of the `WorksIn` relation connects a `Person` node to a `Company` node. In this way all foreign keys can be implemented by reference columns.

Reference columns can also be used to access tables that do not have a primary key. The reference value will be known at the time the referenced row is created and can be used in referring locations or provided by some use-case specific computation, such as MATCH.

Multi-column keys then naturally may be subject to constraints (e.g., “this group of one or more columns in A gives a unique reference value for table A”), and, if keys are used, a foreign key in table B referencing A will have a corresponding group of columns in B.

Many property graph DBMS allow edges to have properties, and in that case an implementation will probably have a table for each edge type to give the properties of each edge and two reference values for the source and destination of the edge. Many SQL tables have several foreign keys, while each edge has only two. Often it will make sense to select different pairs of keys to construct alternative edge models.

The direction of the edge is a property of the edge type rather than its rows (source column TO destination columns), so that if we want to specify this direction in the database, we need the DBMS to support a suitable metadata annotation.

Property graph systems generally use the terminology “node type”, “edge type” both when edges can have their own properties and when they do not. If the details of nodes and edges are saved in a relational database, we should observe that tables and user-defined types are different sorts of database objects in SQL. The values of user-defined types tend to be volatile, and all attributes are usually required to be present and non-null (as in POINT(x,y)). Persistent values of user-defined types can best be placed in a table, and for simplicity in this paper we assume this is happening. The keywords UNDER and ONLY in SQL suggest the possibility of subtype and supertype relationships, and user-defined types can be used in a sort of object-oriented programming.

## III. Ease of Use

Immediate agreement on a common syntax unifying SQL and GQL along these lines is unlikely. Our research supports both, while adding metadata extensions in both SQL and GQL. Noting that the GQL standard does not support an automatic index for a property called ID, we can add SQL-like PRIMARY KEY metadata to element type specifications. To SQL we can add GQL-like connecting metadata to table (and user defined type) creation syntax. To combine named properties with arrow syntax we can add direction to SQL and arrow names to GQL. Let us call such a unified relational system DBR.

Then assuming the current user is allowed to make changes to the schema, consider the effect of a statement such as:

```
INSERT (:Person{name:'Fred'})-worksin
        ->(:Company{name:'AWS'})                        (1)
```

This is not a GQL insert statement as the worksin identifier is just naming the arrow. It could create a `Person` record (with a record identity say f) and a `Company` record (say c), creating new tables (node types) for them, and a reference property `worksin` in Person whose value is the AWS record in Company:

```
 PERSON f { Name:Fred, worksin: Company_c}              (2)
 COMPANY c {Name:AWS}
```

On the other hand, the GQL insert syntax

```
INSERT (:Person{name:'Fred'})-[:WorksIn]
->(:Company{name:'AWS'}))                               (3)
```

mandates a separate edge type `WorksIn` that connects `PERSON` to `COMPANY` (which in GQL is the more natural model). This makes more sense if the edge has its own properties, as in a different example:

```
INSERT (Person{Name:'Mary'})-[:Married '2025-07-24']
      ->(:Person{name:'Joe'})                           (4)
```

Ignore syntax (3) for now and consider implementing (4). If Person Joe is already defined, in GQL (which does not yet have a `MERGE` statement) we would need to retrieve a reference for him first, and write something like

```
MATCH (j:Person{Name:'Joe'}) INSERT
   (:Person{name:'Mary')                                (5)
   -[:Married {Date:2025-07-24}]->(j)
```

Here `Joe` is already constructed: we want to construct a new node for Mary and connect a `Married` edge to `Joe`. The `Match` statement binds j to the `Person` node whose name is `Joe`, so that the insert statement can construct an edge to the new Person node Mary. Such an edge type is declared in GQL by the element specification

```
Directed edge MARRIED {date :: Date} connecting (PERSON
-> PERSON)                                              (6)
```

The SQL equivalent syntax (also supported in DBR) creates a user defined type with the edge connection information (from/to/with and named arrows) as metadata

```
Create type MARRIED as (date Date) edge type (from
bride=PERSON to groom=PERSON)                           (7)
```

The resulting DBR database so far has three tables:

```
Company: (Name string)
Person: (Name string, Worksin ref Company)              (8)
Married: (Bride ref Person, Groom ref Person,
Date date}
```

If `Likes` is implemented as in (1) we might expect the `Person` table to have a `Likes` column of type `set ref Person`.

The key to implementing `MATCH` and `INSERT` is a system index that contains all references (as a mapping of referenced record to property uid and referencing record) on the understanding that each record type maintains an index of its records.

## IV. Supporting Graph Insert and Match Operations

Graph insertion is almost trivial once the identity of any referenced rows is established, and many products offer syntax similar to (1) above. The simple syntax of (1) above can be easily extended to allow comma-separated lists, node references allowing a node or edge constructed in one part of the insert statement to be referred to later.

For pattern matching, the syntax tends to be more complex, but the basic idea can be seen from the example (supposing many records of marriages) :

```
MATCH (p)-[:Married]->(:person{name:'Joe'})         (9)
```

This should return a table of results showing all of the `Persons` P that married a `Person` named `Joe`. In more complex cases, there will be several columns of results and ways of aggregating results (count, sum etc). To implement `MATCH` it is best to work along the `MATCH` expressions identifying what possibilities there are at every stage and backtracking to consider alternatives. The programming technique of continuations can be used for complex patterns where groups of edges can be repeated [14].

Reverse references are best managed by indexes constructed by the DBMS when records are added, and then such indexes can simply use the identity of the new record instead of its values. This infrastructure can be added to any relational database and then used to consider what the graph looks like: we consider such choices in the next section.

Graph models are useful to indicate chains of links (such as transfers of funds, supply chains, etc), and the same business database should be able to support different graph models to allow the study of different sorts of chains. Some graph database management systems implement `MATCH` statements to traverse such chains without constructing joins of tables. A unified implementation of SQL and GQL would allow graph insert and match query by graph patterns, which in some use cases results in large efficiency gains.

## V. Use Role Metadata to Create Type Graph Models

In many use cases, the data in graph models comes from a relational database. Commonly, the node types and edge types in the database have a simple relationship with entities and tables in the relational database. As long as this relationship continues, we expect that execution of a property graph data-modifying statement will make appropriate changes in the underlying relational tables.

In many cases, the graph model comes first (as in (1) above), and the tables in the relational database directly support correspondingly named graph element types (as in (2)). Single reference properties, such as `worksin` implement simple edges without properties.

It is also common for graph models to be built from tables already in a relational database, and in these cases, choices are available. Any relational table can be treated as a node type, and some or all of its reference columns as simple edges. Any relational table with two reference columns can be treated as an edge type, and there is a choice of direction. If a relational table has more than two reference columns then any two columns can become the source and destination columns to make it an edge type. These choices can be saved in graph metadata such as `EDGETYPE(bride=person TO groom=person)`.

Let us briefly consider the `Worksin` and `Married` examples in this context. Suppose the database has been constructed with the three tables as above. The obvious graph model over this seems to have two node types and one edge type. But let us specify that DBR will support a model that also treats Worksin fully as an edge type even though it does not have its own records.

Equally, a graph model that had no interest in dates of marriages could in DBR treat `Married` as a simple column similar to `Worksin`.

Many SQL implementations support the use of roles to identify aliases and privileges for particular classes of users, so it is natural to expect that alternative graph models can be defined for different roles. Many real relational databases are very large, and such a role definition would help to simplify the database for its users by hiding tables and columns that are not relevant to the model, and/or limiting the ability of the role to modify data. While a graph model is being developed, it would be very useful for such role definitions to be automatically generated by inferring the metadata from graph insert patterns such as (1) or (5).

We propose to go further and implement union types and subtypes in our new implementation DBR. Subtype inheritance is already under discussion in the GQL community: multiple node labels (the GQL notion of label sets) can be implemented with union types.

For example, in the GDC Financial benchmark [10] we see edges with union connectors:

```
directed edge Apply {createTime::timestamp,
organization::string} connecting (Person|Company to
Loan)                                            (10)
```

DBR implements this by allowing connector references in column lists in SQL insert statements:

```
    insert into apply (from person, loan, createTime,
organization)
values('57978','4612532092624966603','2020-01-04
22:30:59.239','OneMain Financial')               (11)
```

This mechanism is used when loading the data in the Financial Benchmark from spreadsheet files. Once the data is loaded, the use of rowids instead of key values promises gains in performance.

## VI. An experimental implementation of DBR

An implementation of the above concept is underway, as a research project extending the Pyrrho DBMS [11]. It already has been used to construct the Graph Data Council Financial Benchmark [10]. Current progress already indicates that this implementation behaves as it should when given ordinary relational tasks, atomic, consistent, isolated, and durable (ACID) transactions, procedures, methods, triggers etc.) and offers property graph syntax compatible with the new GQL standard [9].

With the above approach, a referencing property/column can be (a) implied by `FOREIGN KEY` metadata based on key values or (b) defined with `REFERENCES` metadata specifying that its value will identify a row or set of rows in a given target. In both cases the column defines a directed or undirected (possibly set-valued) edge connector. The referent(s) may be of a single table or type, or of a join or disjoint sum of tables/types specified by label set(s) (as in (10) above).

Noting that both SQL and GQL specifications tolerate implementation-specific extensions, the conclusion of the research is that a single implementation of both languages is indeed possible and provides a backward-compatible upgrade path for all existing implementations should they choose to follow it.

Performance and memory usage in the experimental version of Pyrrho DBMS is similar to previous versions: database size on disk is somewhat reduced, use of computer memory slightly increased. It is for the beholder to judge matters such as conceptual clarity and elegance: we do claim that in addition to providing graph facilities to SQL and relational storage to GQL, the proposals here provide a way of clarifying the relationship between the design of a database and the construction of a graph model. Our approach allows a given relational database to support different graph models, offering a choice depending on different use cases.

## VII. Conclusions and Future Work

Even without node and edge types, it is convenient to be able to use graphical insert and MATCH when working with relational databases, to specify MATCH for table constraints, and to define subtypes using MATCH. Importantly, tests include verification that database objects and metadata can be added or dropped during transactional operation.

The accompanying conference presentation will contain examples additional to those above, based on the GDC Financial Benchmark [10], and a tutorial session at DBKDA is planned.

It is hoped to have the implementation of the combined system by the conference date. A tutorial will include hands-on demonstrations.

The Appendix gives a brief test of the current state of implementation, showing the construction of the financial benchmark [10], using the smallest scale factor sf001. The execution time was 49 seconds on  a desktop PC with Ultra i5 procesor.

## APPENDIX

1. The GQL syntax description of the financial benchmark as specified in [10]. This text can be processed by the current version of DBR.

```
create schema /ldbc
[create graph type /ldbc/finBenchMark  {
node Person {id::int,name::string,isBlocked::boolean,
createTime::timestamp,gender::string,birthday:: date,country::
string,city::string},
node Account {id::int,createTime::timestamp,isBlocked::boolean,
type::string,nickname::string,phoneNumber::string,email::string,
freqLoginType::string,lastLoginTime::timestamp,accountLevel:: string},
node Medium {id::int,type::string,isBlocked::boolean,
createTime::timestamp,lastLoginTime::timestamp,riskLevel::string},
node Company{id::int,name::string,isBlocked::boolean,
createTime::timestamp,country::string,city::string,
business::string,description::string, url::string},
node Loan {id::int,loanAmount::float64,balance::float64,
createTime::timestamp,usage::string,interestRate::float32},
directed edge Transfer {amount::float64,createTime::timestamp,
ordernumber::string,comment::string,payType::string,
goodsType::string} connecting (Account to Account),
directed edge Withdraw {createTime::timestamp,amount::float64}
connecting (Account to Account),
directed edge Repay {createTime::timestamp,amount::float64}
connecting (Account to Loan),
directed edge Deposit {createTime::timestamp,amount::float64}
connecting (Loan to Account),
directed edge SignIn {createTime::timestamp,location::string}
connecting (Medium to Account),
directed edge Invest {createTime::timestamp,ratio::float64}
connecting (Person|Company to Company),
directed edge Apply {createTime::timestamp,organization::string}
connecting (Person|Company to Loan),
directed edge Guarantee {createTime::timestamp,relationship::string}
connecting (Person|Company to Person|Company),
directed edge Own {createTime::timestamp}
connecting (Person|Company to Account)}]
```

2. The DBR syntax description of the financial benchmark modified to use inheritance.

```
create type legalentity as(name string,isBlocked boolean,createtime
timestamp,country string,city string) nodetype
create type person under legalentity as (id int primary key,gender
string,birthday timestamp)
create type company under legalentity as(cid int primary key,business
string,description string,url string)
create type actbase as(createtime timestamp,accttype string) nodetype
alter table actbase add balance float64 default 0.0
create type account under actbase as(id int primary key,isBlocked
boolean,nickname string,phonenumber string,email string,freqlogintype
string,lastlogintime timestamp,accntlevel string)
create type loan under actbase as (lid int primary key,loanAmt
float64,interest float32)
create type medium as(id int primary key,type string,isblocked
boolean,createtime timestamp,lastlogin timestamp,risklevel string)
nodetype
create type apply as (createTime timestamp,organization string)
edgetype(legalentity,loan)
create type personapplyloan under apply edgetype(person,loan)
create type companyapplyloan under apply edgetype(company,loan)
create type guarantee as (createTime timestamp,relationship string)
edgetype (legalentity,legalentity)
create type personguaranteeperson under guarantee
edgetype(person,person)
create type invest as (ratio float64,createTime timestamp)
edgetype(legalentity,legalentity)
create type personinvestcompany under invest edgetype(person,company)
create type owns as (createTime timestamp)
edgetype(legalentity,actbase)
create type personownsaccount under owns edgetype(person,account)
create type activatedfor as (createtime timestamp,location string)
edgetype (medium,account)
create type transfer as(amount float64,createTime
timestamp,orderNumber int,comment string,payType string,goodsType
string) edgetype(from actbase with activatedfor optional to actbase)
create type accountrepayloan under transfer edgetype(from account with
activatedfor optional to loan)
create type accounttransferaccount under transfer edgetype(from
account with activatedfor optional to account)
create type companyguaranteecompany under guarantee
edgetype(company,company)
create type companyinvestcompany under invest
edgetype(company,company)
create type companyownsaccount under owns edgetype(company,account)
create type loandepositaccount under transfer edgetype(from loan with
activatedfor optional to account)
create type mediumsigninaccount as (createtime timestamp, location
string) edgetype(medium,account)
```

3. The construction of sf001 with the above DBR specification (recorded on 13 February 2026).

```
|----------------------|
|CURRENT_TIME          |
|----------------------|
|739659.15:20:10.5268673|
|----------------------|
QL> insert into person(id,name,isBlocked,createTime,gender,birthday,country,city) values ~c:\LDBC\sf001\Pe
son.csv
785 records affected in sf001
QL> insert into account(id,createTime,isBlocked,acctType,nickname,phonenumber,email,freqLogintype,lastLogi
time,accntlevel) values ~c:\LDBC\sf001\Account.csv
2055 records affected in sf001
QL> insert into company(cid,name,isBlocked,createTime,country,city,business,description,url) values ~c:\LD
C\sf001\Company.csv
386 records affected in sf001
QL> insert into loan(lid,loanamt,balance,createTime,accttype,interest) values ~c:\LDBC\sf001\Loan.csv
1376 records affected in sf001
QL> insert into medium(id,type,isblocked,createtime,lastlogin,risklevel) values ~c:\LDBC\sf001\Medium.csv
978 records affected in sf001
QL> set referencing // transform ID references into positions
QL> insert into personapplyloan(from1,to1,createTime,organization) values ~c:\LDBC\sf001\PersonApplyLoan.c
v
927 records affected in sf001
QL> insert into personguaranteeperson(from1,to1,createTime,relationship) values ~C:\LDBC\sf001\PersonGuara
teePerson.csv
377 records affected in sf001
QL> insert into personinvestcompany(from1,to1,ratio,createTime) values ~c:\LDBC\sf001\PersonInvestCompany.
sv
1304 records affected in sf001
QL> insert into personownsaccount(from1,to1,createTime) values ~c:\LDBC\sf001\PersonOwnAccount.csv
1384 records affected in sf001
QL> insert into accountrepayloan(from1,to1,amount,createTime) values ~c:\LDBC\sf001\AccountRepayLoan.csv
2747 records affected in sf001
QL> insert into accounttransferaccount(from1,to1,amount,createTime,orderNumber,comment,payType,goodsType)
alues ~c:\LDBC\sf001\AccountTransferAccount.csv
8132 records affected in sf001
QL> insert into accounttransferaccount(to1,from1,amount,createTime) values ~c:\LDBC\sf001\AccountWithdrawA
count.csv
9182 records affected in sf001
QL> insert into companyapplyloan(from1,to1,createtime,organization) values ~c:\LDBC\sf001\CompanyApplyLoan
csv
449 records affected in sf001
QL> insert into companyguaranteecompany(from1,to1,createTime,relationship) values ~c:\LDBC\sf001\CompanyGu
ranteeCompany.csv
202 records affected in sf001
QL> insert into companyinvestcompany(from1,to1,ratio,createTime) values ~c:\LDBC\sf001\CompanyInvestCompan
.csv
679 records affected in sf001
QL> insert into companyownsaccount(from1,to1,createTime) values ~c:\LDBC\sf001\CompanyOwnAccount.csv
671 records affected in sf001
QL> insert into loandepositaccount(from1,to1,amount,createTime) values ~c:\LDBC\sf001\LoanDepositAccount.c
v
2758 records affected in sf001
QL> insert into mediumsigninaccount(from1,to1,createTime,location) values ~c:\LDBC\sf001\MediumSignInAccou
t.csv
2489 records affected in sf001
QL> select current_time
|----------------------|
|CURRENT_TIME          |
|----------------------|
|739659.15:20:59.1600798|
|----------------------|
```

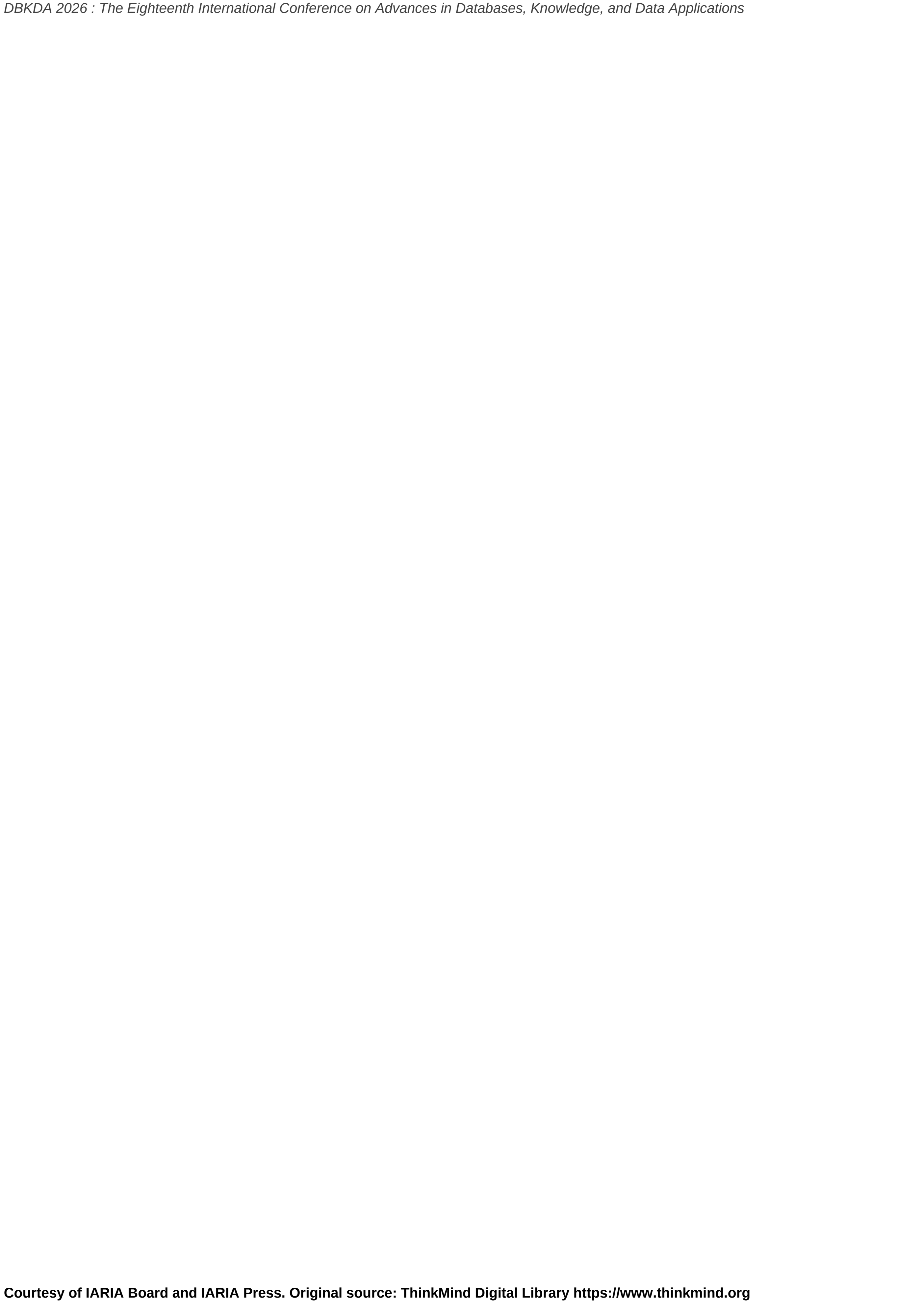